\documentclass[prl,twocolumn,showpacs,preprintnumbers,amsmath,amssymb,floatfix]{revtex4}

\usepackage[]{graphicx}         
\usepackage{bm}             
\usepackage{tabularx}

\newcommand{\degree}{^{\circ}}

\begin{document}

%
%
\title{Generation linewidth of an auto-oscillator with a nonlinear frequency shift: Spin-torque nano-oscillator}

\author{Joo-Von Kim}
\email{joo-von.kim@ief.u-psud.fr}
\affiliation{Institut d'Electronique Fondamentale, UMR CNRS 8622, Universit{\'e} Paris-Sud, 91405 Orsay cedex, France}

\author{Vasil Tiberkevich and Andrei N. Slavin}
\affiliation{Department of Physics, Oakland University, Rochester, MI 48309, USA}

\date{\today}                                           

%
%
\begin{abstract}
It is shown that the generation linewidth of an auto-oscillator with a nonlinear frequency shift (i.e. an auto-oscillator in which frequency depends on the oscillation amplitude) is substantially larger than the linewidth of a conventional quasi-linear auto-oscillator due to the renormalization of the phase noise caused by the nonlinearity of the oscillation frequency. The developed theory, when applied to a spin-torque nano-contact auto-oscillator, predicts a minimum of the generation linewidth when the nano-contact is magnetized at a critical angle to its plane, corresponding to the minimum nonlinear frequency shift, in good agreement with recent experiments.
\end{abstract}


\maketitle

%
%

It is well-known that the linewidth $\Gamma_0$ of a passive oscillating circuit is determined by the ratio of its dissipative element (e.g. resistance $R$) to its reactive element (e.g. inductance $L$): $\Gamma_0 = R/2L$. When the oscillating circuit is connected to an active element (transistor, vacuum tube, tunnel
diode, etc.) and a source of a constant voltage (e.g. battery) the auto-generation of constant-amplitude oscillations at the resonance frequency of the oscillating circuit ($\omega = 1/\sqrt{LC}$, where
$C$ is the circuit capacitance) can take place \cite{Blaquiere, Rabinovich}. The equilibrium amplitude of these auto-oscillations is determined by the dynamic balance between the positive nonlinear
damping of the oscillating system and negative nonlinear damping introduced into the system by the active element \cite{Blaquiere, Rabinovich}.

It is also well-established that the generation linewidth $\Delta\omega$ in a typical auto-oscillator is determined, for the most part, by thermal phase noise (see e.g. Eq.~(9.36) in \cite{Blaquiere}), and can be expressed in the following general form,
\begin{equation}\label{AO-linewidth}
    \Delta\omega = \Gamma_0 \frac{k_{\rm B}T}{E(a)}
\,,\end{equation}
where $k_{\rm B}$ is the Boltzmann constant, $T$ is the absolute temperature, $E(a) = \beta |a|^2$ is the averaged energy of the auto-oscillation having the complex amplitude $a$, and $\beta$ is the coefficient relating the averaged energy to the auto-oscillation amplitude. For example, in an auto-oscillator with a standard linear oscillating circuit, $\beta  = C/2$, where $C$ is the capacitance of the oscillating circuit and $a$ is the amplitude of the voltage on this capacitance. Eq.~(\ref{AO-linewidth}) is rather general and is equally applicable to any type of conventional auto-oscillator (transistor, vacuum tube, tunnel diode, laser, etc.) in which the oscillation frequency is not strongly dependent on the auto-oscillation amplitude, i.e. in the limit $d\omega/d|a|^2 \to 0$.

There exist, however, auto-oscillators for which the oscillation frequency exhibit a strong nonlinearity $N \equiv d\omega/d|a|^2$ that is too large to be neglected. In such systems, one expects that even small fluctuations in the amplitude at steady state can give important contributions to the phase noise. A pertinent example of present interest is the magnetic spin-torque (ST) nano-oscillator~\cite{slonczewski96,berger96,tsoi98,kiselev03,rippard04a}, which consists of a nano-sized metallic contact attached to a magnetic multilayer. Direct electrical current passing through the nano-contact can lead to a transfer of spin-angular momentum between magnetic layers in the stack~\cite{slonczewski96,berger96}, which in turn creates an effective negative damping for the magnetization of the thinner ("free") magnetic layer. This negative damping, analogous to the role played by an active element, can lead to  self-sustained oscillations of magnetization in the free layer. The frequency of these auto-oscillations is determined by the applied magnetic field, static magnetization, etc., and is, in general, close to the ferromagnetic resonance (FMR) frequency in this layer.

In contrast to traditional quasi-linear auto-oscillators, the frequency of the ST nano-oscillator strongly depends on the amplitude of the magnetization precession $a$: $\omega(a) = \omega_0 + N|a|^2$. The sign and magnitude of the nonlinear frequency shift coefficient $N$ depend on the direction and magnitude of the bias magnetic field (see~\cite{kiselev03,rippard04a,rippard04b,slavin05a} for details), and can be varied over range comparable to the oscillation frequency itself. Thus, the classical result (\ref{AO-linewidth}) cannot describe quantitatively the generation linewidth in ST oscillators, and a new theory that explicitly takes into account the nonlinear frequency shift of the auto-oscillator is necessary.

In this Letter, we develop a theory to describe the generation linewidth in an auto-oscillator with nonlinear frequency shifts, and show that this nonlinearity leads to significant linewidth broadening. The theory is then applied to the ST nano-oscillator and we demonstrate that the correct treatment of such nonlinearities is essential for even the \emph{qualitative} description of the nonlinear auto-oscillator.

It has been shown previously~\cite{slavin05a,rezende05a,phaselock06} that the dynamics of the dimensionless complex amplitude $a$ (normalized by $|a|^2=(M_0-M_z)/2M_0$, where $M_0$ is the length of the magnetization vector in the free magnetic layer, and $M_z$ is the projection of this vector on the equilibrium magnetization direction $\bf z$; see \cite{slavin05a} for details), which describes magnetization precession in an ST oscillator, is given by
\begin{eqnarray}\label{model}
    \frac{\partial a}{\partial t} +i(\omega_0 + N |a|^2)a + \Gamma_0(1+Q_0|a|^2)a
&&\\\nonumber
                - \Gamma_s(1-Q_s|a|^2)a &=& f_n(t)
\,.\end{eqnarray}
$\omega_0$ is the linear oscillation frequency and $N$ is the nonlinear frequency shift coefficient (see Eq.~(38) in \cite{slavin05a}). $\Gamma_0$ is the linear part of the positive damping and characterizes the equilibrium linewidth in the passive regime, and $Q_0$ is a phenomenological coefficient characterizing the nonlinearity of positive damping (see~\cite{NonlinearGilbert} for details). $\Gamma_s$ is an effective negative damping introduced into the oscillating system by the active element (in the case of a spin-torque oscillator $\Gamma_s = \sigma I$, where $I$ is the bias current and $\sigma$ is the spin-polarization efficiency defined in Eq.~(2) of \cite{phaselock06}), and the coefficient $Q_s$ describes the nonlinearity of the negative damping ($Q_s = 1$ for ST oscillator \cite{slavin05a}). $f_n(t)$ is a stochastic term that accounts for the influence of the thermal fluctuations (noise). The function $f_n(t)$ is a white Gaussian noise with zero mean and second-order correlator
\begin{equation}\label{noise}
    \langle f_n(t)f_n^*(t') \rangle = 2\Gamma_0 P_n \delta(t-t')
\ .\end{equation}
Here $P_n$ is the oscillator power at thermal equilibrium, i.e. $\langle |a|^2 \rangle_{\Gamma_s = 0} = P_n$. Note that the oscillator energy can be written as $E(a) = \beta |a|^2$, where the constant $\beta$ depends on the normalization of the oscillator amplitude $a$. Then, the equilibrium noise power $P_n$ can be written as $P_n = k_{\rm B}T/\beta$, where $k_{\rm B}$ is the Boltzmann constant and $T$ is the absolute temperature of the system.

We would like to stress that while Eq.~(\ref{model}) is obtained for the case of a spin-torque oscillator, it can adequately describe an auto-oscillator of {\em any} nature under the influence of white noise $f_n(t)$, provided that this oscillator has a nonlinear frequency dependence $(\omega_0 + N|a|^2)$, nonlinear natural positive damping $\Gamma_0(1+Q_0|a|^2)$,
and nonlinear negative damping $\Gamma_s(1-Q_s|a|^2)$.

The stationary solution of Eq.~(\ref{model}) in the absence of noise ($f_n(t) = 0$) can be easily
obtained,
\begin{equation}\label{no-noise}
    a(t) = \sqrt{P_0}e^{-i\omega t + i\phi}
\,,\end{equation}
where
\begin{eqnarray}
    P_0 &=& \frac{\zeta-1}{Q_s\zeta+Q_0}
\,,\\
    \omega &=& \omega_0 + N P_0 = \omega_0 + N\frac{\zeta-1}{Q_s\zeta+Q_0}
\,.\end{eqnarray}
$\phi$ is a constant oscillation phase and $\zeta \equiv \Gamma_s/\Gamma_0$ is the supercriticality parameter. The
solution Eq.~(\ref{no-noise}) exists only for $\zeta > 1$.

Sufficiently far above the threshold (i.e. for $P_0 \gg P_n$) the solution of Eq.~(\ref{model}) with the noise term included will be similar to the noise-free solution Eq.~(\ref{no-noise}) in the sense that the oscillation amplitude will be close to the mean value of $\sqrt{P_0}$, i.e. $|a(t)| = \sqrt{P_0} + \delta A(t)$, $|\delta A(t)|^2 \ll P_0$, and the phase $\phi$ will be a slow function of time. Substituting the expression
\begin{equation}\label{anzats}
    a(t) = \left(\sqrt{P_0} + \delta A(t)\right)e^{-i\omega t + i\phi(t)}
\end{equation}
for $a(t)$ in Eq.~(\ref{model}), and retaining only the terms of the first order in $\delta A$, we find equations for the amplitude fluctuations,
\begin{equation}\label{eq:linearampfluc}
    \frac{\partial \delta A}{\partial t} +2\Gamma_{\rm eff}P_0\delta A = {\rm Re}(\tilde f_n(t)e^{-i\phi}),
\end{equation}
and phase fluctuations,
\begin{equation}\label{eq:linearphasefluc}
    \frac{\partial \phi}{\partial t} +2N\sqrt{P_0}\delta A = \frac{1}{\sqrt{P_0}}\,{\rm Im}(\tilde f_n(t)e^{-i\phi})
\ .\end{equation}
Here the effective damping $\Gamma_{\rm eff}$ is expressed as
\begin{equation}
    \Gamma_{\rm eff} = \Gamma_0 \, (Q_s\zeta+Q_0),
\end{equation}
and $\tilde f_n(t) = f_n(t)e^{i\omega t}$. Note, that the statistical properties of $f_n(t)$ and $\tilde f_n(t)$ are identical. Therefore, tilde will be omitted in the following text for simplicity.

There is a significant qualitative difference between the behavior of the amplitude fluctuations described by
Eq.~(\ref{eq:linearampfluc}) and the behavior of the phase fluctuations described by Eq.~(\ref{eq:linearphasefluc}). Since the oscillation amplitude at steady state remains practically constant on average in the presence of thermal fluctuations, $|a| \approx\sqrt{P_0}$, the correlation function for the amplitude fluctuations $K_A(\tau) \equiv \langle |a(t)||a(t+\tau)| \rangle$ remains finite even if $\tau \to \infty$, i.e. $K_A \to P_0$. Therefore, for large $\tau$ the behavior of the full correlation function $K(\tau) \equiv
\langle a(t)a^*(t+\tau) \rangle$ will be determined solely by the phase fluctuations,
\begin{equation}
    K(\tau) \approx P_0 \langle e^{i[\phi(t)-\phi(t+\tau)]} \rangle e^{i\omega\tau}
\ .\end{equation}

For the frequency linewidth of the auto-oscillation, we are interested only in the fluctuations taking place inside a narrow frequency region $\Delta\omega \ll \Gamma_{\rm eff}P_0$, in which $|\partial\delta A/\partial t| \sim \Delta\omega|\delta A| \ll 2\Gamma_{\rm eff}P_0|\delta A|$. As such, the first (derivative) term in the left hand side of Eq.~(\ref{eq:linearampfluc}) can be neglected compared to the second term, and an explicit expression for $\delta A(t)$ can be obtained,
\begin{equation}
    \delta A = \frac{1}{2\Gamma_{\rm eff}P_0}\,{\rm Re}(f_ne^{-i\phi})
\ .\end{equation}
Substituting this expression for $\delta A(t)$ in Eq.~(\ref{eq:linearphasefluc}) leads to a closed-form equation for the phase fluctuations $\phi(t)$ in the system,
\begin{eqnarray}\label{eq:phi}
    \frac{\partial \phi}{\partial t} &=& \frac{1}{\sqrt{P_0}}\left[
                - \frac{N}{\Gamma_{\rm eff}}\,{\rm Re}(f_ne^{-i\phi})
                + \,{\rm Im}(f_ne^{-i\phi})
              \right],
\nonumber\\&=&
              \frac{1}{\sqrt{P_0}}\sqrt{1+\left(\frac{N}{\Gamma_{\rm eff}}\right)^2}\,{\rm Im}(f_ne^{-i\alpha-i\phi})
\,,\end{eqnarray}
where $\alpha = {\rm atan}(N/\Gamma_{\rm eff})$.

Eq.~(\ref{eq:phi}) is formally identical to the equation for phase fluctuations in a system {\em without a nonlinear frequency shift} (see e.g. second equation (9.8) in \cite{Blaquiere}), but with the \emph{increased noise level}
\begin{equation}\label{renormalization}
    f_n(t) \to f'_{n}(t) = \sqrt{1+\left(\frac{N}{\Gamma_{\rm eff}}\right)^2}e^{-i\alpha}f_n(t)
\ .\end{equation}
Applying the general methodology to compute oscillator linewidths in the \textit{absence of nonlinear frequency shifts} (see, e.g., Chapter 9 in \cite{Blaquiere} or \cite{kim06}) leads to an expression for the Lorentzian linewidth of the auto-oscillator \textit{with a nonlinear frequency shift} $N$,
\begin{equation}\label{linewidth1}
    \Delta\omega = \Gamma_0\left(\frac{k_{\rm B}T}{E_0}\right)\left[1+\left(\frac{N}{\Gamma_{\rm eff}}\right)^2\right]
\ .\end{equation} We have rewritten the ratio $P_n/P_0$ as $k_{\rm
B}T/E_0$, where $E_0= \langle E(a)\rangle =\beta P_0$ is the average
oscillator energy. A comparison of the classical result
(\ref{AO-linewidth}) with the generalized Eq.~(\ref{linewidth1})
shows clearly that the nonlinear frequency shift in the
auto-oscillator leads to a significant linewidth broadening that is
due to an effective renormalization of the phase noise
(\ref{renormalization}) by the nonlinearity $N$.

The result in (\ref{linewidth1}) is the principal result of this paper and illustrates the fact that three key parameters determine the linewidth of an auto-oscillator \textit{with a nonlinear frequency shift}. First, the equilibrium relaxation rate of the oscillator $\Gamma_0$ determines the overall scale of the possible linewidth variations. Second, the generation linewidth is proportional to the ratio of the noise energy (which increases with temperature) to the average energy of the auto-oscillation. Third, the ratio of the nonlinear frequency shift coefficient $N$ to the effective nonlinear damping $\Gamma_{\rm eff}$ gives a measure of the phase noise renormalization due to amplitude fluctuations.

For the particular case of an ST nano-oscillator an attempt to
calculate the generation linewidth was previously undertaken by one
of the authors in \cite{kim06}, but the nonlinear frequency shift
was neglected. The calculation resulted in an expression for the
linewidth (see Eq.~(28) in \cite{kim06}) that can be cast in the
classical form (\ref{AO-linewidth}), where the constant $\beta$ is
given by $\beta = (M_0/\gamma) \omega_0 V_{\rm eff}$, where $\gamma$
is the gyromagnetic ratio, $\omega_0$ is the oscillation frequency,
and $V_{\rm eff}$ is the effective volume of the magnetic material
of the free layer involved in the auto-oscillation.

Note that the effective volume $V_{\rm eff}$ for a ST auto-oscillator based on a magnetic nanopillar is equal to the volume of the nanopillar free layer itself, $V_{\rm eff} = V$, which is
substantially smaller than the effective volume for a similarly-sized magnetic nano-contact. In the latter the effective volume is defined by Eq.~(4) in \cite{phaselock06} and  is several times larger than the physical volume of the free magnetic layer \textit{under the contact} due to the propagation of spin waves radiated from the nano-contact. Thus, our theory finds a natural explanation for a well-known experimental fact (see~\cite{kiselev03,rippard04a, rippard04b}) that the auto-oscillation linewidths associated with magnetic nanopillars are, in general, several times broader that those in magnetic nano-contacts.

Another important result that follows from Eq.~(\ref{linewidth1}), in the context of a ST auto-oscillator, is the prediction of a linewidth minimum that follows from a change in sign in the frequency shift (e.g., from ``red" ($N < 0$) to ``blue" ($N > 0$)) as the magnetization is tilted out of the film plane. Across this transition the nonlinear frequency shift coefficient $N$ passes through zero (see, e.g., Fig.~8 in \cite{slavin05a}) at which one recovers the smaller value of the phase linewidth. A change in magnetization angle can be achieved in practice by applying a large external magnetic field at different orientations out of the film plane. This linewidth minimum has been recently observed in experiment (see Fig.~6 in \cite{rippard06a}).

To illustrate the qualitative behavior of the generation linewidth of a ST auto-oscillator with the variation of external control parameters, such as the magnitude of the bias current $I$ (which is proportional to the effective negative damping $\Gamma_s \equiv \sigma I$ in Eq.~(\ref{model})) and the magnitude $H_e$ and direction $\theta_e$  of the external bias magnetic field, we present below results of the linewidth obtained from Eq.~(\ref{linewidth1}). We consider a circular nanopillar oscillator with the following parameters of the free layer: saturation magnetization $\mu_0 M_0 = 1$~T, Gilbert damping constant $\alpha_G = 0.01$, nonlinear damping coefficient $Q_0 = 0$, thickness $L = 5$~nm, and diameter $D = 50$~nm. For simplicity we assume that spin-polarized current excites spin wave mode with a frequency close to the FMR frequency in a continuous free layer (see Eq.~(37) in \cite{slavin05a}). In that case the natural linear damping $\Gamma_0$ and the coefficient of the nonlinear frequency shift $N$ are given by Eq.~(31) and Eq.~(38) of \cite{slavin05a}, respectively.

\begin{figure}
\includegraphics[width=0.45\textwidth]{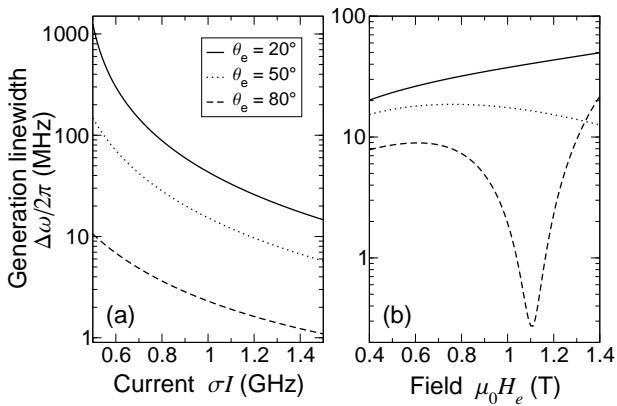}
\caption{\label{figure01} Generation linewidth  of a spin-torque auto-oscillator calculated
from Eq.~(\ref{linewidth1}) as a function of: (a) negative current-induced damping $\Gamma_s\equiv\sigma I$ for $\mu_0 H_e = 1.2$~T; (b) applied bias magnetic field $H_e$ for $\sigma I = 1.0$~GHz; for several angles $\theta_e$ between the bias field and the plane of the free layer.}
\end{figure}

\begin{figure}
\includegraphics[width=0.375\textwidth]{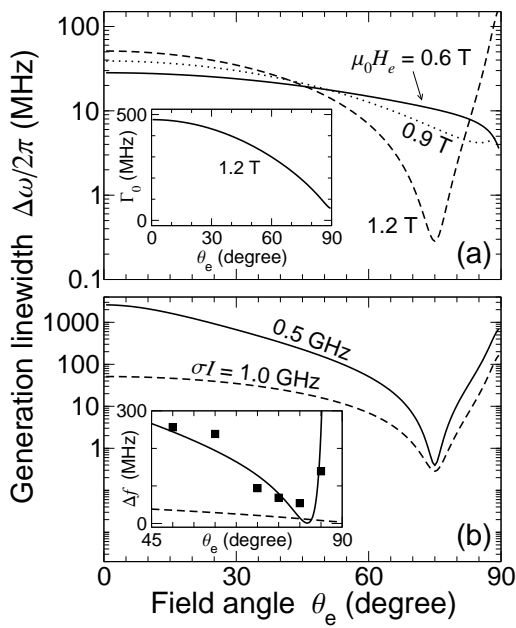}
\caption{\label{figure02}
Generation linewidth as a function of applied field angle $\theta_e$ for: (a) three applied fields at constant $\sigma I = 1$~GHz; (b) two bias currents at constant $\mu_0 H_e = 1.2$~T. Inset of (a): Equilibrium linewidth $\Gamma_0$ as a function of $\theta_e$ for $ \mu_0 H_e = 1.2$~T. Inset of (b): Comparison of $\Delta f = \Delta \omega / 2\pi$ from Eq.~(\ref{linewidth1}) (solid line), $10\times \Delta f$ from Eq.~(\ref{AO-linewidth}) (dashed line) with $I = 9$~mA and $\mu_0H_e = 0.9$~T, and experimental data from Fig.~6a in Ref.~\onlinecite{rippard06a} (squares).
}
\end{figure}

While linewidth narrowing is expected with the increase in bias current due to the increase in auto-oscillation amplitude (Fig.~\ref{figure01}a), the linewidth behavior as a function of the bias magnetic field is more complex than what might be expected from (\ref{linewidth1}). Indeed, one observes a non-monotonous variation in the linewidth around field magnitude and orientations that lead to a change in sign in $N$. In such regions, there is a rapid linewidth narrowing followed by a broadening as the field is increased through the critical region in which the linewidth minimum occurs ($\theta_e = 80\degree$ in Fig.~\ref{figure01}b).
This complex linewidth behavior further is exemplified in Fig.~\ref{figure02} in which we present the role of the applied field orientation. The position of the linewidth minimum is clearly sensitive to the field magnitude and orientation, as it is the combination of these parameters that determine the three critical parameters in (\ref{linewidth1}), namely, $\Gamma_0$, $\Gamma_{\rm eff}$, and $N$.
This linewidth dependence on the field orientation is in good qualitative agreement with the experimental results from Ref.~\onlinecite{rippard06a}. A quantitative comparison between the theory and experiment is shown in the inset of Fig.~\ref{figure02}b in which the generation linewidth was calculated for the experimental nanocontact geometry with the same parameters as above except for $\mu_0M_0 = 0.75$~T and $Q = 3$. In contrast to the classical result (\ref{AO-linewidth}), which predicts much narrower lines and a monotonous decrease in the linewidth as a function of $\theta_e$, the renormalized phase-noise result (\ref{linewidth1}) gives a reasonable qualitative and quantitative description of the experimentally observed behavior, in particular, the linewidth minimum around $\theta_e \approx 80\degree$.

In summary, we have developed a theory of the generation linewidth of a nonlinear auto-oscillator with a nonlinear frequency shift which generalizes the classical result (\ref{AO-linewidth}). The additional nonlinearity in the oscillator frequency leads to a renormalization of the phase noise far above threshold. Applied to the particular case of a spin-torque nano-oscillator, the theory accounts for a number of characteristic, but previously unexplained, features observed in experiment: (i) General linewidth narrowing with increases in the bias current and oscillation amplitude (see Fig.~4 in Ref.~\onlinecite{rippard06a}); (ii) Presence of a linewidth minimum as a function of the external magnetic field orientation (see Fig.~6 in Ref.~\onlinecite{rippard06a}); (iii) Narrower lines in nanocontacts compared to nanopillars.

\begin{acknowledgments}
This work was in part supported by the MURI grant W911NF-04-1-0247 from the Department of Defense of the USA, by the grant W911NF-04-1-0299 from the U.S. Army Research Office, and by the Oakland University Foundation. JK acknowledges support from the European Communities program IST under Contract No. IST-016939 TUNAMOS.
\end{acknowledgments}

\clearpage


\begin{thebibliography}{99}

\bibitem{Blaquiere}
A. Blaquiere,
\textit{Nonlinear System Analysis}
(Academic Press, N.Y., 1966).

\bibitem{Rabinovich}
M. I. Rabinovich and D.I. Trubetskov, \textit{Oscillations and Waves
in Linear and Nonlinear Systems} (Kluwer, Dordrecht-Boston, 1989).

\bibitem{slonczewski96}
J. C. Slonczewski,
J. Magn. Magn. Mat. {\bf 159}, L1 (1996).

\bibitem{berger96}
L. Berger,
Phys. Rev. B {\bf 54}, 9353 (1996).

\bibitem{tsoi98}
M. Tsoi {\it et al.},
Phys. Rev. Lett. {\bf 80}, 4281 (1998).

\bibitem{kiselev03}
S. I. Kiselev {\it et al.},
Nature (London) {\bf 425}, 380 (2003).

\bibitem{rippard04a}
W. H. Rippard {\it et al.},
Phys. Rev. Lett. {\bf 92}, 027201 (2004).

\bibitem{rippard04b}
W. H. Rippard {\it et al.}, Phys. Rev. B {\bf 70}, 100406(R) (2004).

\bibitem{slavin05a}
A. N. Slavin and P. Kabos,
IEEE Trans. Magn. {\bf 41}, 1264 (2005).

\bibitem{rezende05a}
S. M. Rezende, F. M. de Aguiar, and A. Azevedo, Phys. Rev. Lett. {\bf 94}, 037202 (2005).

\bibitem{phaselock06}
A. N.~Slavin  and V.S.~Tiberkevich, Phys. Rev. B {\bf 74}, 104401 (2006).

\bibitem{NonlinearGilbert}
V. S. Tiberkevich and A. N. Slavin, 
Phys. Rev. B {\bf 75}, 014440 (2007).

\bibitem{kim06}
J.-V. Kim, Phys. Rev. B {\bf 73}, 174412 (2006).

\bibitem{rippard06a}
W. H.~Rippard {\it et al.}, 
Phys. Rev. B {\bf 74}, 224409 (2006).

\end{thebibliography}
\end{document}